\documentclass[aps,prb,twocolumn,nobibnotes]{revtex4-2}  
\usepackage{dcolumn}
\usepackage{graphicx}
\usepackage{color}
\usepackage{amssymb}
\usepackage{amsmath}
\usepackage{bm}
\usepackage[colorlinks=true,linktocpage=true,breaklinks=true]{hyperref}
\usepackage{multirow}
\usepackage{braket}
\usepackage{appendix}

\begin{document}

\title{Spin current leakage and Onsager reciprocity in interfacial spin-charge interconversion}

\author{Aur\'{e}lien Manchon$^1$}
\email{email: aurelien.manchon@univ-amu.fr}
\author{Shuyuan Shi$^{2,3}$}
\author{Hyunsoo Yang$^2$}
\affiliation{$^1$Aix-Marseille Universit\'e, CNRS, CINaM, Marseille, France}
\affiliation{$^2$Department of Electrical and Computer Engineering, National University of Singapore, 117576, Singapore}
\affiliation{$^3$Fert Beijing Institute, MIIT Key Laboratory of Spintronics, School of Integrated Circuit Science and Engineering, Beihang University, Beijing 100191, China.}
\date{\today}

\begin{abstract}
Experimental investigations of spin-charge interconversion in magnetic bilayers comprising a ferromagnet adjacent to a topological insulator have reported scattered results on the spin-charge and charge-spin conversion efficiency. Attempting to reconcile these contradicting experimental results, we develop a phenomenological theory of spin-charge interconversion accounting for both interfacial interconversion through the spin galvanic effect, also called the Rashba-Edelstein effect, as well as bulk interconversion via the spin Hall effect. We find that the spin current leakage into the nonmagnetic metal plays a central role during the spin-to-charge and charge-to-spin conversion, leading to dissymmetric interconversion processes. In particular, spin-to-charge conversion is much less sensitive to the spin current absorption in the nonmagnetic metal than charge-to-spin conversion. This suggests that spin pumping is a more trustable technique to extract the interfacial Rashba parameter than spin-orbit torque.  
\end{abstract}

\maketitle

\section{Introduction} 
Spin-to-charge interconversion mediated by spin-orbit coupling has become a central technique in spintronics enabling the electrical manipulation of magnetization \cite{Manchon2019}, as well as the generation of charge currents induced by magnetization precession \cite{Han2018,Nan2021,Kondou2023}. The standard system is composed of a magnetic thin film deposited on top of a nonmagnetic metal with strong spin-orbit coupling. Whereas the original workhorse of spin-charge interconversion was NiFe/Pt \cite{Saitoh2006}, a broad range of materials has been explored over the years, including both metallic \cite{Mosendz2010,Rojas-Sanchez2014,Pham2016} and insulating ferromagnets \cite{Heinrich2011,Hahn2013,Hahn2013b,Avci2017}, antiferromagnets \cite{Vaidya2020,Li2020c}, adjacent to strongly spin-orbit coupled materials including transition metals (Pt, W, and Ta). In these systems, the interconversion is mostly attributed to the spin Hall effect (SHE) taking place in the bulk of the heavy metal \cite{Hoffmann2013b,Sinova2015}. In the past decade, the attention has been shifted towards materials displaying large interfacial spin-momentum locking. Recent research encompasses metals with strong interfacial Rashba states \cite{Rojas-Sanchez2013b,Zhang2015a,Karube2016} (Ag/Bi, Ag/Sb or Ag/Bi$_{2}$O$_{3}$) , topological insulators \cite{Mellnik2014,Fan2014a,Deorani2014,Shiomi2014,Wang2015b,Jamali2015,Wang2016l,Kondou2016,Rojas-Sanchez2016b,Wang2017d,Mahendra2018,Wu2019c,Bonell2020,He2021b} (Bi$_2$Se$_3$, etc.), transition metal dichalcogenides \cite{Shao2016,MacNeill2017,Shi2019,Xue2020,Xie2021} (MoS$_2$, WTe$_2$), van der waals materials \cite{SaveroTorres2017,Safeer2019,Benitez2020,Shin2022,Xie2022b} (for non-local interconversion, see Refs. \cite{Choi2022,Ontoso2023}), oxide heterostructures \cite{Lesne2016,Wang2017e,Vaz2019,Trier2020,Gallego2023} (see Ref. \cite{Trier2022} for a review), chiral metals \cite{Calavalle2022} and the so-called ferroelectric Rashba semiconductors such as SrTiO$_3$ and $\alpha$-GeTe \cite{Noel2020,Varotto2022b}, whose interfacial spin-momentum locking can be controlled by switching the ferroelectric polarization \cite{Rinaldi2018}. 

In all these materials, strong spin-orbit coupling and inversion symmetry breaking at the interface result in Dirac or Rashba spin-momentum locking \cite{Bihlmayer2022}, which enables interconversion between charge and spin currents via the so-called (inverse) spin galvanic effect (SGE), also known as the Rashba-Edelstein effect \cite{Ivchenko1978,Vasko1979,Edelstein1990}. In contrast, the SHE that converts three-dimensional (3D) spin currents into 3D charge currents, the interfacial spin galvanic effect converts a two-dimensional (2D) charge current into a 3D spin current, and vice-versa. The interconversion efficiency of SHE is quantified by the dimensionless parameters $\xi_{\rm cs}$ and $\xi_{\rm sc}$. In a bulk heavy metal, Onsager reciprocity dictates that $\xi_{\rm cs}=\xi_{\rm sc}$. As discussed in more detail in the present work, in the case of a magnetic bilayer, this equality holds true as long as the charge current density involved in the definition of these efficiencies is restricted to the charge current flowing in the heavy metal.

The interconversion efficiency of SGE is rather quantified in terms of charge-to-spin conversion length $\lambda_{\rm sc}$ and spin-to-charge conversion inverse length $q_{\rm cs}$. From a theoretical standpoint, $\xi$ is related to the SHE in the bulk heavy metal modulated by the spin transparency, i.e., the ability of the heavy metal to absorb the spin current. The figures of merit $\lambda_{\rm sc}$ and $q_{\rm cs}$ have been recently derived for a 2D Rashba or Dirac gas in Refs. \cite{Zhang2016n,Dey2018,Isshiki2020,He2021b} using either phenomenological or microscopic approaches. These references provide the general form $\lambda_{\rm sc}=(\alpha_{\rm R}/\hbar)(1/\tau_p+1/\tau_t)$ and $q_{\rm cs}=(\alpha_{\rm R}/\hbar v_F^2)\tau_t$, where $\tau_t$ is the tunneling time between the 2D gas and the spin current source (typically an adjacent ferromagnet, but also possibly a nonmagnetic metal through which the spin current is injected) and $\tau_p$ is the momentum relaxation time in the 2D gas. When emphasize that the validity of these theories, as well as the very definition of $\lambda_{\rm sc}$ and $q_{\rm cs}$, is restricted to the ideal case of 2D transport. In other words, whether these parameters are meaningful in the case of oxide heterostructures where the charge current is purely two-dimensional, it may simply not apply in other metallic systems or in weakly insulating "topological insulators" such as Bi$_2$Se$_3$.

As a matter of fact, experiments in topological insulators report large variations in spin-charge interconversion efficiency. In particular, it is often observed that spin-to-charge and charge-to-spin conversion efficiencies massively differ from each other. Charge-to-spin conversion in Bi-based topological materials ranges from 0.4 to more than 400 \cite{Mellnik2014,Fan2014a,Shiomi2014,Wang2015b,Kondou2016,Wang2017d,Mahendra2018,Wu2019c} whereas spin-to-charge conversion rather ranges from 0.0001 to 0.4
\cite{Deorani2014,Shiomi2014,Jamali2015,Wang2016l}, which cannot be properly accounted for using the simple formulas given above. This set of results contrasts sharply with the metallic case, where the reported efficiencies, $\xi_{\rm cs}$ and $\xi_{\rm sc}$, are usually of the same order. 

In an attempt to clarify the situation, we develop a phenomenological model for spin-charge interconversion in a metallic bilayer where interfacial SGE and bulk SHE coexist. We show that whereas spin-to-charge and charge-to-spin conversion are intrinsically asymmetric, although not breaking Onsager reciprocity, the leakage of the spin current in the neighboring nonmagnetic metal plays a crucial role when the conductivity of the nonmagnetic metal becomes large. 

\begin{figure}[ht!]
\includegraphics[width=1\linewidth]{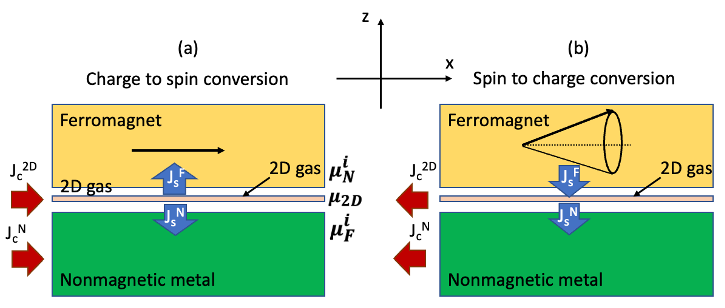}
\caption{(Color online). (a) Charge-to-spin and (b) spin-to-charge conversion in a metallic bilayer composed of a 2D gas embedded between a ferromagnet and a nonmagnetic metal. }
\label{fig0}
\end{figure} 

\section{Drift-diffusion theory} 
We consider the system represented in Fig. \ref{fig0} and composed of three elements: a 2D gas embedded between a ferromagnet F and a nonmagnetic metal N, of thicknesses $d_{\rm F}$ and $d_{\rm N}$, respectively. The 2D gas can be thought of as the surface state of a heavy metal (e.g., Co/Pt, Ag/Bi) or that of a weakly conductive topological insulator (e.g., Bi$_2$Se$_3$). For simplicity, we assume that the spin-charge conversion in the 2D gas takes place through the SGE (or Rashba-Edelstein) effect, with an effective coupling $\alpha_R$ (eV.m), and in the nonmagnetic metal, it is induced by SHE with an angle $\theta_H$ (unitless).
\begin{eqnarray}
{\bf J}_c^{\rm 2D}=\sigma_{\rm 2D}{\bf E}-e^2{\cal N}_{\rm 2D}\frac{\alpha_R}{\hbar}{\bf z}\times{\bm \mu}_{\rm 2D},
\end{eqnarray}
where the second term is the SGE that converts a non-equilibrium spin density into a charge current. Here, ${\bm\mu}_{\rm 2D}=\mu_{\rm 2D} {\bf y}$ (in the units of V) is the spin chemical potential, ${\cal N}_{\rm 2D}$ (${ eV^{-1}\cdot m^{2}}$) and $\sigma_{\rm 2D}$ ($\Omega^{-1}$) are the density of states and conductivity of the 2D gas. The spin continuity equation reads
\begin{eqnarray}
\partial_t{\bm\mu}_{\rm 2D}=\frac{\alpha_R}{\hbar}{\bf z}\times{\bf E}-\frac{{\bm\mu}_{\rm 2D}}{\tau_{\rm sf}}-\frac{{\bm\mu}_{\rm 2D}-{\bm\mu}_{\rm F}^i}{\tau_{\rm F}}-\frac{{\bm\mu}_{\rm 2D}-{\bm\mu}_{\rm N}^i}{\tau_{\rm N}}.\label{eq:2}
\end{eqnarray}
The first term is the inverse SGE, i.e., the generation of a spin density induced by the flow charge current, with $\alpha_{\rm R}$ being the so-called Rasbha parameter. A formal derivation of the spin diffusion equation in a Rashba gas can be found in Refs. \onlinecite{Mishchenko2004,Burkov2004,Wang2012b}. The second term accounts for the spin relaxation in the gas (that may be anisotropic, although for the sake of simplicity this is neglected in the present study) and the last two terms account for spin injection in the neighboring layers: $1/\tau_{{\rm F(N)}}$ is the tunneling rate between the 2D gas and the ferromagnet (nonmagnetic metal), and ${\bm\mu}_{{\rm F(N)}}^i$ is the interfacial spin chemical potential in the ferromagnet (nonmagnetic metal). These tunneling rates are related to the interfacial spin currents flowing between the 2D gas and the ferromagnetic (${\bf J}_s^{{\rm F},i}$) and nonmagnetic metals (${\bf J}_s^{{\rm N},i}$) through the boundary conditions
\begin{eqnarray}
{\bf J}_s^{{\rm N},i}=G_{\rm N}({\bm\mu}_{\rm N}^i-{\bm\mu}_{\rm 2D}),
\end{eqnarray}
where $G_{\rm N}=e^2{\cal N}_{\rm 2D}/\tau_{\rm N}$ ($\Omega^{-1}\cdot{\rm m}^{-2}$) is the interfacial conductance between the 2D gas. Notice that without loss of generality, the spin current is expressed in the same units as the charge current for simplicity. Similarly,
\begin{eqnarray}\label{eq:4}
{\bf J}_s^{{\rm F},i}=2G_{\uparrow\downarrow}{\bm\mu}_{\rm 2D}-{\bf J}_s^0,
\end{eqnarray}
where ${\bf J}_s^0$ is due to spin pumping (if any), and $2G_{\uparrow\downarrow}=e^2 {\cal N}_{\rm 2D}/\tau_F$ is the spin mixing conductance  ($\Omega^{-1}\cdot{\rm m}^{-2}$), assuming that the magnetization is perpendicular to the spin density in the 2D gas (which is valid in both spin-to-charge and charge-to-spin interconversion scenarios), i.e., ${\mu}_{F}^i=0$. 

In the nonmagnetic metal, the (3D) charge and spin currents, ${\bf J}_c^{\rm N}$ and ${\bf J}_s^{\rm N}$, are defined
\begin{eqnarray}
{\bf J}_c^{\rm N}&=&\sigma_{\rm N}{\bf E}+\theta_H\sigma_{\rm N}{\bm\nabla}\times{\bm \mu}_{\rm N},\\
{\bf J}_s^{\rm N}&=&-\sigma_{\rm N}{\bm\nabla}{\bm \mu}_{\rm N}-\theta_H\sigma_{\rm N}{\bf y}\times{\bf E}.
\end{eqnarray}
Here, one recognizes the SHE and inverse SHE, $\sigma_{\rm N}$  ($\Omega^{-1}\cdot{\rm m}^{-1}$) being the conductivity of the metal. In the scenario studied in the present work, the spin polarization remains aligned along {\bf y} and therefore all the equations above can be simply projected along this direction. Applying the two boundary conditions $J_s^{\rm N} (z=0)=G_{\rm N} (\mu_{\rm N}^i-\mu_{\rm 2D})$ and $J_s^{\rm N} (z=-d_{\rm N} )=0$, we obtain 
\begin{eqnarray}\label{eq:spinacc}
\mu_{\rm N} (z)&=&\frac{1}{1+\eta_{\rm N}}\left(\eta_{\rm N}\mu_{\rm 2D}+\tilde{\lambda}_{\rm N}\tilde{\theta}_H E\right)\frac{\cosh\frac{z+d_{\rm N}}{\lambda_{\rm N}}}{\cosh\frac{d_{\rm N}}{\lambda_{\rm N}}}\nonumber\\
&&+\tilde{\lambda}_{\rm N}\tilde{\theta}_H E\frac{\tanh\frac{d_{\rm N}}{\lambda_{\rm N}}\sinh\frac{z}{\lambda_{\rm N}}}{\cosh\frac{d_{\rm N}}{\lambda_{\rm N}}-1},\end{eqnarray}

where $\tilde{\theta}_H=\theta_H\left(1-\cosh^{-1}\frac{d_{\rm N}}{\lambda_{\rm N}}\right)$, $\tilde{\lambda}_{\rm N}=\lambda_{\rm N}/\tanh\frac{d_{\rm N}}{\lambda_{\rm N}}$, and $\eta_{\rm N}=\tilde{\lambda}_{\rm N}G_{\rm N}/\sigma_{\rm N}$. This relation does not depend on the scenario (spin-to-charge or charge-to-spin conversion) and shows that the nonmagnetic metal always acts like a spin sink since (for $E=0$), $\eta_{\rm N}\rightarrow+\infty\Rightarrow\mu_{\rm N}^i=\mu_{\rm 2D}$ (perfect spin insulator) and $\eta_{\rm N}\rightarrow0\Rightarrow\mu_{\rm N}^i=0$ (perfect spin sink). Let us now compute the spin-charge interconversion efficiency in the cases depicted in Fig. \ref{fig0}.

\subsection{Charge to spin conversion}

An electric field ${\bf E}$ is applied along ${\bf x}$ and generates a spin accumulation, polarized along ${\bf y}$, throughout the structure [see Eq. \eqref{eq:spinacc}]. This spin accumulation injects a spin current $J_s^{{\rm F},i}$ into the ferromagnet. In the absence of spin pumping, $J_s^{{\rm F},i}=2G_{\uparrow\downarrow}\mu_{\rm 2D}$ [Eq. \eqref{eq:4}]. The spin chemical potential $\mu_{\rm 2D}$ can be obtained from the spin continuity equation, Eq. \ref{eq:2}, posing $\partial_t\mu_{\rm 2D}=0$ and injecting $\mu_{\rm N}^i=\mu_{\rm N} (0)$ [Eq. \eqref{eq:spinacc}]. We get
\begin{eqnarray}
\mu_{\rm 2D}&=&\frac{\frac{\alpha_R}{\hbar}+\frac{1}{\tau_{\rm N}}\frac{\tilde{\lambda}_{\rm N}\tilde{\theta}_H}{1+\eta_{\rm N}}}{\frac{1}{\tau_{\rm sf}} +\frac{1}{\tau_{\rm F}}+\frac{1}{\tau_{\rm N}}\frac{1}{1+\eta_{\rm N}}}E,
\end{eqnarray}
which implies that
\begin{eqnarray}
J_s^{{\rm F},i}&=&2G_{\uparrow\downarrow}\frac{\frac{\alpha_R}{\hbar}+\frac{1}{\tau_{\rm N}}\frac{\tilde{\lambda}_{\rm N}\tilde{\theta}_H}{1+\eta_{\rm N}}}{\frac{1}{\tau_{\rm sf}} +\frac{1}{\tau_{\rm F}}+\frac{1}{\tau_{\rm N}}\frac{1}{1+\eta_{\rm N}}}E.
\end{eqnarray}
To obtain, the unitless charge-to-spin interconversion efficiency, we divide the spin current by the current flowing in both the nonmagnetic metal and 2D gas, $J_c=\frac{\sigma_{\rm N}d_{\rm N}+\sigma_{\rm 2D}}{d_{\rm N}+t_{\rm 2D}}E$, where $t_{\rm 2D}$ is the effective thickness of the 2D gas. We then obtain,
\begin{eqnarray}\label{eq:cs}
\xi_{\rm cs}=\left |\frac{J_s^{{\rm F},i}}{J_c}\right |&=&2G_{\uparrow\downarrow}\frac{d_{\rm N}+t_{\rm 2D}}{\sigma_{\rm N}d_{\rm N}+\sigma_{\rm 2D}}\frac{\frac{\alpha_R}{\hbar}+\frac{1}{\tau_{\rm N}}\frac{\tilde{\lambda}_{\rm N}\tilde{\theta}_H}{1+\eta_{\rm N}}}{\frac{1}{\tau_{\rm sf}} +\frac{1}{\tau_{\rm F}}+\frac{1}{\tau_{\rm N}}\frac{1}{1+\eta_{\rm N}}}.
\end{eqnarray}
One immediately notices the competition between the spin relaxation inside the 2D gas ($\sim 1/\tau_{\rm sf}$), the spin absorption in the nonmagnetic metal ($\sim 1/\tau_{\rm N}$) and in the ferromagnet ($\sim 1/\tau_{\rm F}\propto G_{\uparrow\downarrow}$). This competition will be analyzed in more details in the next section. For now, we emphasize that other renormalizations can be adopted. For instance, assuming that the charge-to-spin conversion only occurs in the 2D gas ($\theta_H=0$), one can normalize the spin current to the charge current that effectively flows along the interface, $J_c^{\rm 2D}=\sigma_{\rm 2D}E$, which yields,
\begin{eqnarray}
q_{\rm cs}=\left |\frac{J_s^{{\rm F},i}}{J_c^{\rm 2D}}\right |&=&\frac{\alpha_R}{\hbar}\frac{2G_{\uparrow\downarrow}}{\sigma_{\rm 2D}}\frac{1}{\frac{1}{\tau_{\rm sf}} +\frac{1}{\tau_{\rm F}}+\frac{1}{\tau_{\rm N}}\frac{1}{1+\eta_{\rm N}}}.
\end{eqnarray}
This expression extends the ones derived in Refs. \onlinecite{Zhang2016n,Dey2018,Isshiki2020,He2021b} beyond the model Rashba or Dirac gas, and includes the effect of the adjacent spin sink, providing a more accurate expression for the figure of merit of interfacial spin-to-charge conversion.

\subsection{Spin to charge conversion}

A spin current $J_s^{{\rm F},i}=-(e^2 {\cal N}_{\rm 2D})/\tau_F \mu_{\rm 2D}+J_s^0$  is pumped out of the ferromagnet and converted into a charge current composed of $J_c^{\rm 2D}=e^2 {\cal N}_{\rm 2D} (\alpha_R/\hbar)\mu_{\rm 2D}$ and  $J_c^{\rm N}=-(1/d_{\rm N})\int_{-d_{\rm N}}^0dz\theta_H \sigma_{\rm N} \partial_z\mu_{\rm N}$. The spin continuity equation, Eq. \eqref{eq:2}, gives
\begin{eqnarray}
\mu_{\rm 2D}=\frac{J_s^0}{e^2 {\cal N}_{\rm 2D}}\frac{1}{\frac{1}{\tau_{\rm sf}} +\frac{1}{\tau_{\rm F}}+\frac{1}{\tau_{\rm N}}\frac{1}{1+\eta_{\rm N}}}. 
\end{eqnarray}
Therefore, we deduce the charge current flowing in the 2D gas and in the nonmagnetic metal,
\begin{eqnarray}
J_c^{\rm 2D}&=&\frac{\alpha_R}{\hbar}\frac{1}{\frac{1}{\tau_{\rm sf}} +\frac{1}{\tau_{\rm F}}+\frac{1}{\tau_{\rm N}}\frac{1}{1+\eta_{\rm N}}}J_s^0,\\
J_c^{\rm N}&=&\frac{\tilde{\theta}_H \tilde{\lambda}_{\rm N}}{d_{\rm N}}\frac{1}{\frac{1}{\tau_{\rm sf}} +\frac{1}{\tau_{\rm F}}+\frac{1}{\tau_{\rm N}}\frac{1}{1+\eta_{\rm N}}}J_s^0.
\end{eqnarray}
The total current that is pumped through the system is therefore $I_c^{\rm pump}=w(J_c^{\rm N} d_{\rm N}+J_c^{\rm 2D})$. In other words, if the effective \{nonmagnetic+2D gas\} thickness is $d_{\rm N}+t_{\rm 2D}$, the effective current density pumped out of the system is $J_c^{\rm pump}=(J_c^{\rm N} d_{\rm N}+J_c^{\rm 2D})/(d_{\rm N}+t_{\rm 2D})$. Finally, the spin-to-charge efficiency reads
\begin{eqnarray}\label{eq:sc}
\xi_{\rm sc}=\left|\frac{J_c^{\rm pump}}{J_s^{{\rm F},i}}\right|=\frac{1}{d_{\rm N}+t_{\rm 2D}}\frac{\frac{\alpha_R}{\hbar}+\frac{1}{\tau_{\rm N}}\frac{\tilde{\lambda}_{\rm N}\tilde{\theta}_H}{1+\eta_{\rm N}}}{\frac{1}{\tau_{\rm sf}} +\frac{1}{\tau_{\rm N}}\frac{1}{1+\eta_{\rm N}}}.
\end{eqnarray}
Comparing Eqs. \eqref{eq:cs} and \eqref{eq:sc}, one can notice that - besides an obvious geometrical factor - these expressions differ on the role played by the spin mixing conductance, $G_{\uparrow\downarrow}\propto1/\tau_{\rm F}$, so that in general $\xi_{\rm cs}\neq\xi_{\rm sc}$. Notice though that this inequality does not mean that Onsager reciprocity is broken at all. As a matter of fact, Onsager reciprocity applies to the response tensor of generalized currents to thermodynamical forces, or, in other words, to the connection between current densities and chemical potential gradients (for a discussion on Onsager reciprocity applies to spin pumping and spin torque, see Ref. \onlinecite{Brataas2012c}). Since the spin-charge interconversion efficiencies are defined as the ratio between two current densities, they do not fulfill Onsager reciprocity. 

In fact, two limits illustrate the difference between spin-to-charge and charge-to-spin conversion in metallic bilayers. In the absence of interfacial spin-charge interconversion, $G_{\rm N}\rightarrow+\infty$, $t_{\rm 2D}\rightarrow0$, $\sigma_{\rm 2D}\rightarrow0$ and $\tau_{\rm sf}\rightarrow+\infty$, we obtain
\begin{eqnarray}\label{eq:sc2}
\xi_{\rm sc}=\frac{\tilde{\lambda}_{\rm N}}{d_{\rm N}}\tilde{\theta}_H\approx\xi_{\rm cs}=\tilde{\theta}_H.
\end{eqnarray}
These expressions are valid in the limit $2G_{\uparrow\downarrow}\tilde{\lambda}_{\rm N}/\sigma_{\rm N}\gg1$. 

Finally, assuming only interfacial spin-to-charge conversion is allowed, the charge current is only pumped in the 2D gas and the spin-to-charge conversion length reads
\begin{eqnarray}\label{eq:lsc}
\lambda_{\rm sc}&=&\frac{\alpha_R}{\hbar}\frac{1}{\frac{1}{\tau_{\rm sf}} +\frac{1}{\tau_{\rm N}}\frac{1}{1+\eta_{\rm N}}},
\end{eqnarray}
which qualitatively agrees with the expressions derived in Refs. \onlinecite{Zhang2016n,Dey2018,Isshiki2020,He2021b}. The difference is that in Eq. \eqref{eq:lsc}, the momentum relaxation time $\tau_p$ is replaced by the spin-flip time in the 2D gas $\tau_{\rm sf}$, and the tunneling time $\tau_t$ is replaced by $\tau_{\rm N}$, accounting for the backflow spin current in the nonmagnetic metal. 
\section{Influence of spin current leakage}
\subsection{Phenomenological parameters}

Besides the conventional electronic properties of the metallic layers (conductivity, spin diffusion length, spin Hall angle), the phenomenological theory developed in the previous section is controlled by two parameters: the coupling between the ferromagnetic layer and the 2D gas, $1/\tau_{\rm F}=G_{\uparrow\downarrow}/e^2{\cal N}_F$, and the coupling between the 2D gas and the nonmagnetic metal, $1/\tau_{\rm N}=G_{\rm N}/e^2{\cal N}_N$. Qualitatively, the interfacial areal conductance between two metallic layers is about $e^2/hA\approx10^{14}-10^{15}\;\Omega^{-1}\cdot$m$^{-2}$ (see, e.g., Refs. \onlinecite{Zwierzycki2005,Brataas2006}), $A$ being the area of a unit cell at the interface. Assuming an interfacial density of state of about $1/({\rm eV}\cdot A)$, one arrives at a coupling rate of $1/\tau_{\rm N,F}\approx10^{15}$ s$^{-1}$, which seems reasonable for direct coupling between metallic states and remains one to two orders of magnitude larger than the typical spin relaxation time in metals.

\subsection{Interfacial interconversion}
\begin{figure}[ht!]
\includegraphics[width=\linewidth]{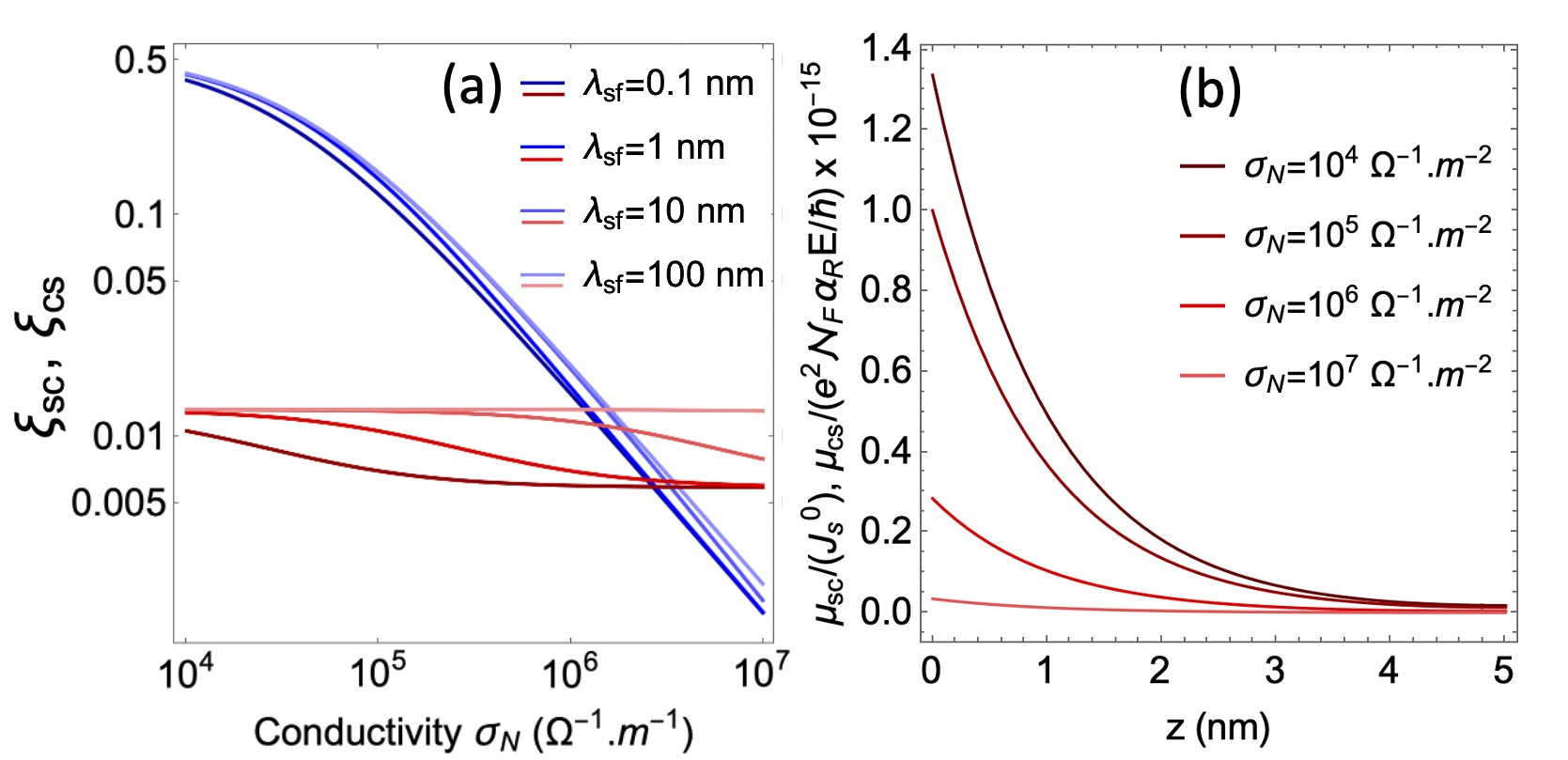}
\caption{(Color online) (a) Spin-to-charge (red) and charge-to-spin (blue) conversion efficiency as a function of the conductivity of the nonmagnetic metal for different values of the spin diffusion length $\lambda_{\rm N}$. (b) Spin accumulation profile in the nonmagnetic metal for different values of the conductivity $\sigma_{\rm N}$. In these calculations, we set $\tau_{\rm sf}=10^{-14}$ s, $\sigma_{\rm 2D}=2\times 10^{-4}$ $\Omega^{-1}$, $d_{\rm N}=5$ nm and $t_{\rm 2D}=1$ nm.\label{fig:fig1}}
\label{fig1}
\end{figure}

Let us first consider the case where the spin-charge interconversion is solely due to spin-momentum locking at the interface. In this case, the adjacent metallic layer does not participate to the conversion itself but only in the spin absorption. To model this case, we set the following parameters, $G_{\rm N}=G_{\uparrow\downarrow}=4\times 10^{14}\;\Omega^{-1}\cdot$m$^{-2}$. In Fig. \ref{fig:fig1}(a), we report the dependence of the interconversion efficiencies, $\xi_{\rm sc}$ (red) and $\xi_{\rm cs}$ (blue), as a function of the nonmagnetic metal conductivity $\sigma_{\rm N}$ for various values of the spin diffusion length $\lambda_{\rm N}$. Notice that the figure is given in logarithmic scale. Clearly, the spin-to-charge and charge-to-spin conversion efficiencies are markedly different in the limit of weakly conducting nonmagnetic metal ($\sigma_{\rm N}\approx 10^4$ $\Omega^{-1}\cdot$m$^{-1}$), providing an explanation for the experimental values reported in bilayers involving topological insulators (Bi$_2$Se$_3$, etc.), as pointed out above. Upon increasing the conductivity of the nonmagnetic metal, the charge-to-spin efficiency (blue) is progressively reduced whereas the spin-to-charge efficiency (red) is only weakly affected. In the case of a conducting nonmagnetic metal ($\sigma_{\rm N}\approx 10^6$ $\Omega^{-1}\cdot$m$^{-1}$), both efficiencies are comparable, as mentioned in the previous section, corroborating the experimental data collected using a conducting heavy metal substrate such as Pt. Reducing the spin diffusion length reduces the interconversion efficiency, as the spin information is lost in the nonmagnetic metal.

In the model proposed above, the interfacial spin chemical potential due to spin pumping, $\mu_{\rm N}^{sc}$, and that due to electrical injection, $\mu_{\rm N}^{cs}$ are related by
\begin{eqnarray}
\frac{\mu_{\rm N}^{sc}(z)}{J_s^0}&=&\frac{\mu_{\rm N}^{cs}(z)}{(e^2{\cal N}_F\alpha_R/\hbar)E},\\
&=&\frac{\eta_{\rm N}}{1+\eta_{\rm N}}\frac{\cosh\frac{z+d_{\rm N}}{\lambda_{\rm N}}}{\cosh\frac{d_{\rm N}}{\lambda_{\rm N}}}\frac{1}{\frac{1}{\tau_{\rm sf}}+\frac{1}{\tau_{\rm F}}+\frac{1}{\tau_{\rm N}}\frac{1}{1+\eta_{\rm N}}}.
\end{eqnarray}
 The profile of the spin chemical potential is reported in Fig. \ref{fig1}(b) for $\lambda_{\rm N}=1$ nm and for various conductivities $\sigma_{\rm N}$. The larger the conductivity the smaller the interfacial spin chemical potential, suggesting that interfacial spin-charge interconversion is enhanced when the adjacent layer is a good spin insulator.

\subsection{Interfacial versus bulk interconversion}

\begin{figure}[ht!]
\includegraphics[width=1\linewidth]{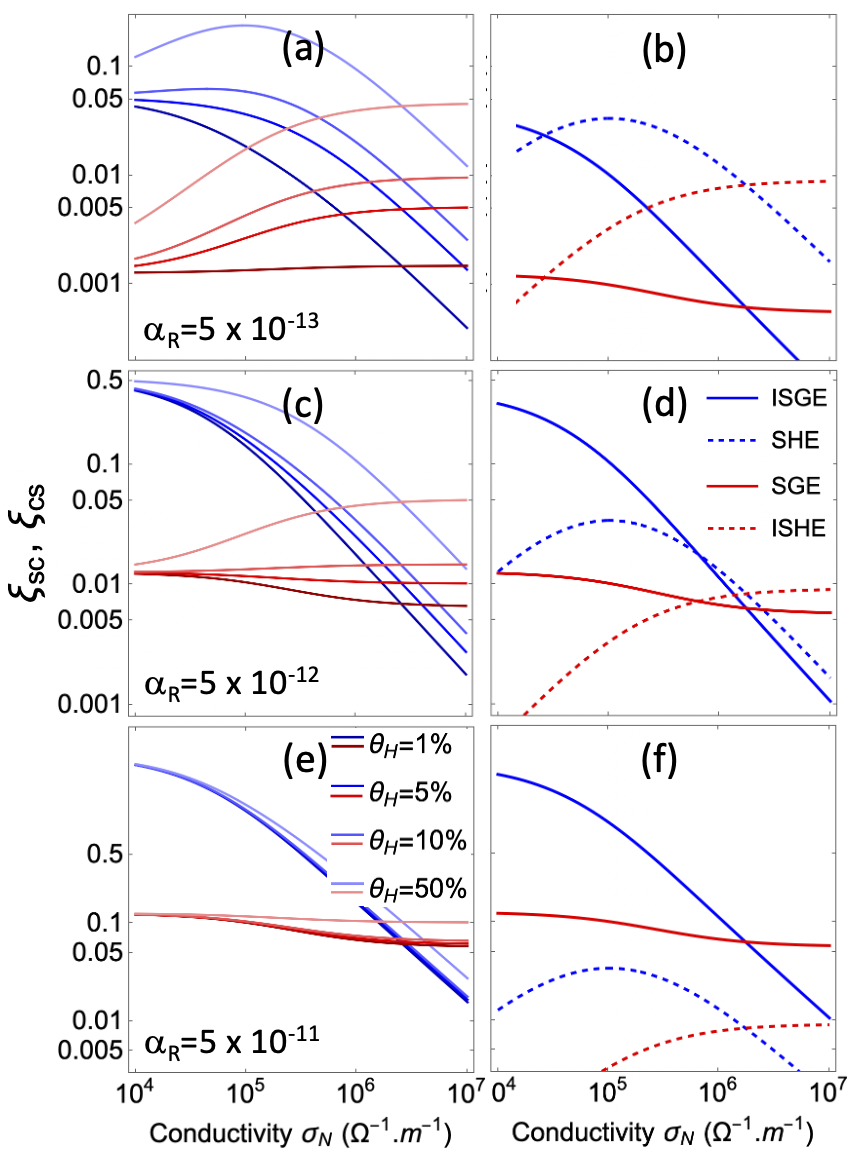}
\caption{(Color online). (Left panels) Spin-to-charge (red) and charge-to-spin (blue) conversion efficiencies for (a) weak, (c) intermediate, and (e) strong interfacial Rashba spin-orbit parameters for various values of the spin Hall angle. (Right panels) Corresponding contributions of the spin Hall effect (SHE), inverse spin Hall effect (ISHE), interfacial spin galvanic effect (SGE), and inverse spin galvanic effect (ISGE), for $\theta_H=10\%$. In all these calculations, we set $\lambda_{\rm N}=1$ nm.}
\label{fig2}
\end{figure}

Let us now turn on the spin Hall angle of the nonmagnetic layer. Figure \ref{fig2} reports the interconversion efficiencies as a function of the nonmagnetic metal conductivity for various magnitudes of the spin Hall angle and for low [Fig. \ref{fig2}(a)], medium [Fig. \ref{fig2}(b)] and strong value of the Rashba term [Fig. \ref{fig2}(e)]. On the right panel, the contribution of SGE (solid) and SHE (dashed)  to spin-to-charge (red) and charge-to-spin conversions (blue). 

In the case of vanishingly weak Rashba strength [Fig. \ref{fig2}(a,b)], the interconversion is dominated by the (inverse) SHE and steadily increases with $\theta_{\rm H}$. The spin-to-charge interconversion efficiency $\xi_{\rm sc}$ (red lines) increases with the conductivity of the normal metal as more output charge current is allowed to flow. On the other hand, the charge-to-spin interconversion efficiency $\xi_{\rm cs}$ (blue lines) decreases in the limit of highly conductive normal metal as the shunting increases. To understand this feature, Fig. \ref{fig2}(b) shows that the contributions of the SGE and ISGE systematically decrease when increasing the conductivity of the nonmagnetic metal, as relatively less current flows inside the 2D gas. Notice that the effect is more dramatic for the spin-to-charge than for the charge-to-spin conversion. The contribution of the SHE is more subtle. Whereas its contribution to the charge-to-spin interconversion increases steadily (dashed red), its contribution to spin-to-charge interconversion first increases, reaches a maximum, and then decreases (dashed blue). This behavior is associated with the competition between the enhanced SHE and the shunting effect.

Upon increasing the strength of the Rashba parameter [Fig. \ref{fig2}(c) and (e)], the spin-to-charge and charge-to-spin conversion efficiencies become less sensitive to the spin diffusion length. Concomitantly, the spin-to-charge conversion (red lines) becomes progressively independent of the nonmagnetic metal conductivity [this is particularly clear in panel (e)], while the charge-to-spin conversion keeps on decreasing. In other words, in the case of dominant interfacial spin-orbit coupling, the spin-to-charge conversion is much less sensitive to the spin current leakage than the charge-to-spin conversion. The decomposition of the conversion efficiencies in terms of SGE and SHE confirms this trend.

\subsection{Extracting interfacial spin-charge interconversion coefficient from experiments}

\begin{figure}[ht!]
\includegraphics[width=1\linewidth]{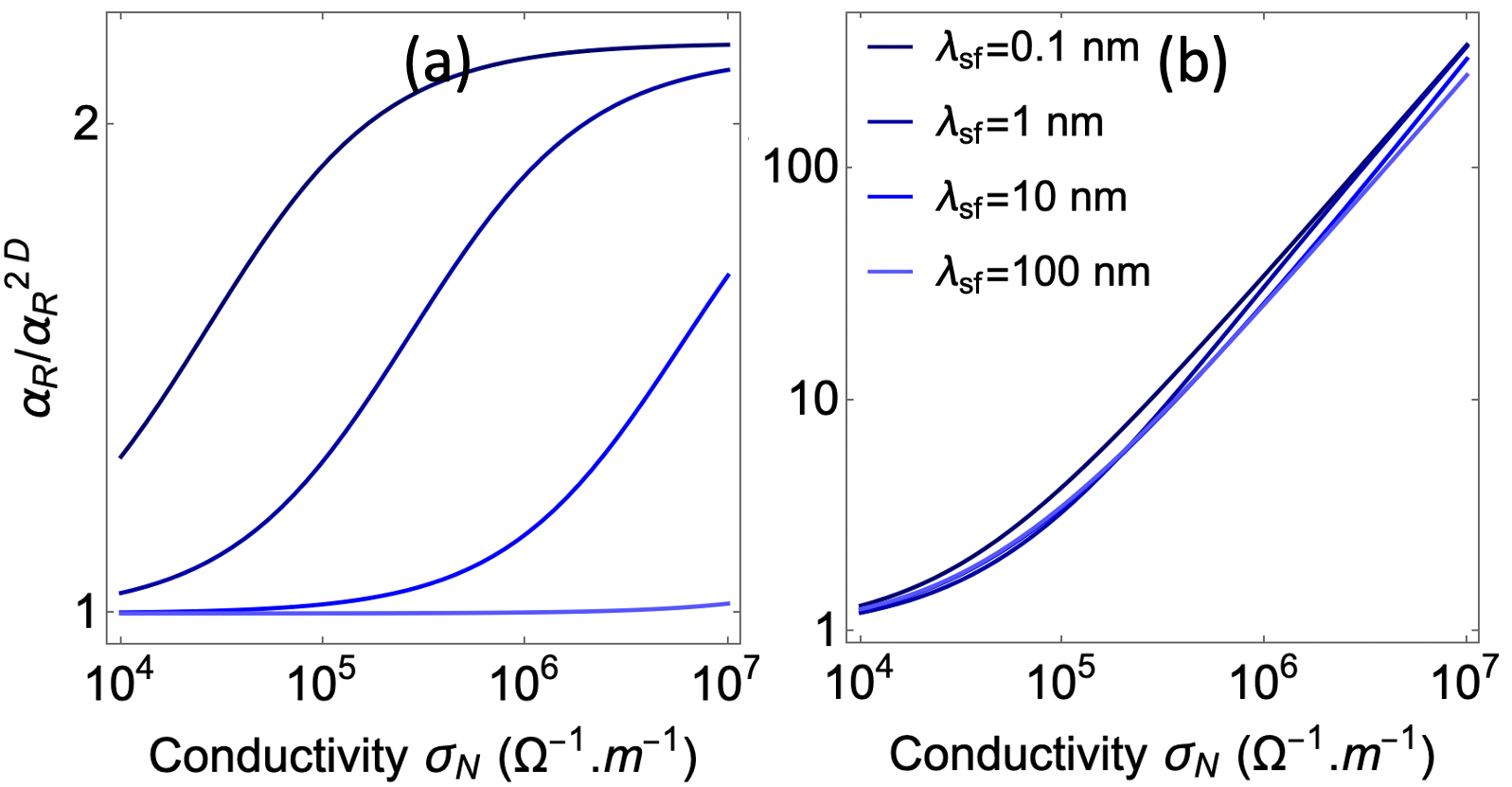}
\caption{(Color online) Ratios between the actual value of the Rashba parameter in a magnetic bilayer and its nominal when only interfacial transport is assumed, for (a) spin-to-charge and (b) charge-to-spin conversion experiments.}
\label{fig3}
\end{figure} 
The spin-charge interconversion coefficients $\xi_{\rm sc}$ and $\xi_{\rm cs}$ are often used to extract the interfacial Rashba parameter, $\alpha_{\rm R}$. The extraction procedure requires the knowledge of basic transport properties such as conductivity, interfacial conductances, and spin diffusion lengths, but also relies on the scenario adopted. Assuming purely 2D transport, the Rashba parameter associated with charge-to-spin conversion reads
\begin{eqnarray}
\frac{\alpha_{\rm R}^{\rm 2D}}{\hbar}=\frac{\xi_{\rm cs}}{2G_{\uparrow\downarrow}}\frac{\sigma_{\rm 2D}}{d_{\rm N}+t_{\rm 2D}}\left(\frac{1}{\tau_{sf}}+\frac{1}{\tau_{F}}\right),
\end{eqnarray}
and the parameter associated with spin-to-charge conversion reads
\begin{eqnarray}
\frac{\alpha_{\rm R}^{\rm 2D}}{\hbar}=\frac{\xi_{\rm sc}}{\tau_{sf}}(d_{\rm N}+t_{\rm 2D}).
\end{eqnarray}
Now, assuming that transport is allowed in the neighboring metal, we obtain
\begin{eqnarray}
\frac{\alpha_{\rm R}}{\hbar}&=&\frac{\xi_{\rm cs}}{2G_{\uparrow\downarrow}}\frac{\sigma_{\rm 2D}+\sigma_{\rm N}d_{\rm N}}{d_{\rm N}+t_{\rm 2D}}\left(\frac{1}{\tau_{sf}}+\frac{1}{\tau_{F}}+\frac{1}{\tau_{\rm N}}\frac{1}{1+\eta_{\rm N}}\right)\nonumber\\
&&-\frac{1}{\tau_{\rm N}}\frac{\tilde{\lambda}_{\rm N}\tilde{\theta}_{\rm N}}{1+\eta_{\rm N}},
\end{eqnarray}
for charge-to-spin conversion and
\begin{eqnarray}
\frac{\alpha_{\rm R}}{\hbar}=\xi_{\rm sc}(d_{\rm N}+t_{\rm 2D})\left(\frac{1}{\tau_{sf}}+\frac{1}{\tau_{\rm N}}\frac{1}{1+\eta_{\rm N}}\right)-\frac{1}{\tau_{\rm N}}\frac{\tilde{\lambda}_{\rm N}\tilde{\theta}_{\rm N}}{1+\eta_{\rm N}},
\end{eqnarray}
for spin-to-charge conversion. To assess how spin current leakage affects the extraction of the interfacial Rashba parameter, Fig. \ref{fig3} shows the ratio $\alpha_{\rm R}/\alpha_{\rm R}^{\rm 2D}$ as a function of the nonmagnetic metal conductivity for (a) spin-to-charge and (b) charge-to-spin conversion experiments. For these simulations, we assumed $\xi_{\rm sc}=\xi_{\rm cs}=50\%$. Clearly, in the limit of an insulating nonmagnetic metal ($\sigma_{\rm N}\rightarrow0$), $\alpha_{\rm R}\rightarrow\alpha_{\rm R}^{\rm 2D}$ and in the limit of highly conducting nonmagnetic metal, the extracted value of $\alpha_{\rm R}$ increases substantially. Interestingly, spin-to-charge and charge-to-spin conversion lead to very different behaviors. When $\xi_{\rm sc}$ is used to quantify the Rashba parameter, the value of $\alpha_{\rm R}$ saturates when $\eta_{\rm N}\approx 1$. The maximum value obtained in (a) is about 
\begin{eqnarray}
\frac{\alpha_{\rm R}}{\alpha_{\rm R}^{\rm 2D}}\rightarrow1+(\tau_{sf}/\tau_{\rm N})\left(1-\frac{\tilde{\lambda}_{\rm N}\tilde{\theta}_{\rm N}}{\xi_{\rm sc}(d_{\rm N}+t_{\rm 2D})}\right),
\end{eqnarray}
which yields a saturation value of the order of 2.2 for our set of parameters. In contrast, using the charge-to-spin conversion efficiency $\xi_{\rm cs}$, the extracted parameter is proportional to the conductivity of the nonmagnetic metal,
\begin{eqnarray}
\frac{\alpha_{\rm R}}{\alpha_{\rm R}^{\rm 2D}}\rightarrow \frac{\sigma_{\rm N}d_{\rm N}}{\sigma_{\rm 2D}}\frac{\frac{1}{\tau_{sf}}+\frac{1}{\tau_{F}}+\frac{1}{\tau_{\rm N}}\frac{1}{1+\eta_{\rm N}}}{\frac{1}{\tau_{sf}}+\frac{1}{\tau_{F}}}.
\end{eqnarray}

Based on these observations, one can draw several conclusions. First, neglecting the potential spin current leakage and its associated SHE can lead to major discrepancies between the extracted interfacial Rashba parameter and the real one. A critical knob in the present theory is the spin transparency $\eta_{\rm N}$. The proper estimate of interfacial spin-charge interconversion requires the accurate determination of this coefficient, i.e., spin relaxation length and interfacial conductance. Finally, it is remarkable that the Rashba parameter extracted from spin-to-charge conversion experiments is much less sensitive to spin current leakage than when using charge-to-spin conversion experiments. This result suggests that spin-pumping is a better tool than spin-orbit torque for the extraction of the interfacial Rashba parameter.

\section{Conclusion}
In order to solve controversial puzzles in spin-charge interconversion experiments, we have developed a phenomenological model that accounts for both interfacial SGE and bulk SHE effect. We find that when interfacial spin-charge interconversion is present, the leakage of spin current into the adjacent nonmagnetic metal substantially affects the overall spin-charge conversion efficiency, even in the absence of SHE. Most interestingly, the spin-to-charge and charge-to-spin conversions are affected differently: converting a charge current into a spin current is much more sensitive to the spin current absorption in the nonmagnetic metal due to the backflow associated with spin relaxation. In contrast, the charge current induced by a spin current is much more robust against spin current leakage. 

This observation opens two interesting avenues. First, it clarifies the recent experimental reports on interconversion in topological insulators and explains why charge-to-spin conversion efficiencies tend to be orders of magnitude larger than spin-to-charge conversion efficiencies \cite{Mellnik2014,Fan2014a,Deorani2014,Shiomi2014,Wang2015b,Jamali2015,Wang2016l,Kondou2016,Rojas-Sanchez2016b,Wang2017d,Mahendra2018,Wu2019c,Bonell2020,He2021b}. In addition, it also indicates that the value of the interfacial Rashba parameter extracted from spin pumping experiments are, a priori, much more trustable than values extracted from spin-orbit torque experiments.

\begin{acknowledgments}
A.M. acknowledges support from the ANR ORION project, Grant No. ANR-20-CE30-0022-01 of the French Agence Nationale de la Recherche as well as the ANR MNEMOSYN project, Grant No. ANR-21-GRF1-0005. S.S. acknowledges support from the National Natural Science Foundation of China, Grant No. 12104032.
\end{acknowledgments}
\bibliography{Biblio2023}

\end{document}